\documentclass[preprint,prd,
amsmath,amssymb,amsthm,nofootinbib]{revtex4}
\usepackage[mathscr]{euscript}
\usepackage{bm}
\usepackage{textcomp}
\usepackage{graphicx}
\usepackage{subfigure}
\usepackage{overpic}
\usepackage{multirow}
\usepackage{floatrow}
\usepackage{float} 
\usepackage{amsmath,amssymb,amsthm}
\usepackage{cases}
\usepackage[colorlinks=true,linkcolor=red]{hyperref}
\newfloatcommand{capbtabbox}{table}[][\FBwidth]
\newcommand{\bea}{\begin{eqnarray}}
\newcommand{\eea}{\end{eqnarray}}
\newcommand{\beq}{\begin{equation}}
\newcommand{\eeq}{\end{equation}}

\def\/{\over}

\begin{document}

\title{Primordial non-Guassianity in inflation with  gravitationally enhanced friction}

\author{ Li-Yang Chen\footnote{clyrion@hunnu.edu.cn}, Hongwei Yu\footnote{
hwyu@hunnu.edu.cn} and Puxun Wu\footnote{
 pxwu@hunnu.edu.cn}  }
\affiliation{Department of Physics and Synergetic Innovation Center for Quantum Effects and Applications, Hunan Normal University, Changsha, Hunan 410081, China 
}
\begin{abstract}
The gravitationally enhanced friction can reduce the speed of the inflaton to realize an ultra-slow-roll inflation, which will amplify the  curvature perturbations. The amplified perturbations  can generate a sizable amount of primordial black holes (PBHs) and induce simultaneously  a significant background gravitational waves (SIGWs). In this paper, we investigate the primordial non-Gaussianity of  the curvature perturbations in the inflation with  gravitationally enhanced friction. We find that when the gravitationally enhanced friction plays a role in the inflationary dynamics, the non-Gaussianity is noticeably larger  than that from the standard slow-roll inflation. During the regime in which  the power spectrum of the curvature perturbations is around its peak,  the non-Gaussianity  parameter changes from negative  to positive. When the power spectrum is at its maximum,  the non-Gaussianity parameter is near zero ($\sim \mathcal{O}(0.01)$). Furthermore,  the primordial non-Gaussianity promotes  the formation of PBHs, while its effect  on SIGWs is negligible.
 \end{abstract}


\maketitle
\section{Introduction}
\label{sec_in}

Inflation resolves most of the problems, such as the flatness, horizon and monopole problems, that plague the standard cosmological model~\cite{Guth1980,Linde1982,Starobinsky1980,Albrecht1982}. During inflation the curvature perturbations are stretched  outside the Hubble horizon and then  stop propagating with the amplitudes frozen at  certain nonzero values. Inflation  predicts a nearly scale-invariant spectrum for the curvature perturbations, which is well consistent with the CMB observations~\cite{Aghanim2020}. The CMB observations indicate  that the amplitude $\mathcal {P_R}$ of the  power spectrum of  the curvature perturbations is about $10^{-9}$~\cite{Aghanim2020}.  After inflation, these super-horizon perturbations, which will reenter the Hubble radius during the radiation- or matter-dominated era, result in the formation of large scale  cosmic structures and at the same time lead to possible generation of primordial black holes (PBHs)~\cite{Hawking1971,Carr1974,Carr1975,Khlopov2007}.  The possibility is however slim for the standard slow-roll inflation since  the amplitude of the power spectrum of the curvature perturbations is too small ($\sim10^{-9}$). 

If  a sizable amount of  PBHs is formed in the early universe,  PBHs with different masses can be used to explain different astronomical  events. For example, the $\mathcal{O}(10) M_\odot$,  $\mathcal{O}(10^{-5}) M_\odot$ and $\mathcal{O}(10^{-12}) M_\odot$ PBHs can explain the gravitational wave events observed by the LIGO/Virgo collaboration~\cite{lg1,lg2,lg3,lg4}  and six  ultrashort-timescale microlensing events in the OGLE data~\cite{P.Mroz2017,H.Niikura2019}, and make up all dark matter~\cite{A.Katz2018,H.Niikura20191,A.Barnacka2012,P.W.Graham2015,Belotsky2014}, respectively, where $M_\odot$ is the mass of the Sun. 
To generate  abundant PBHs,  $\mathcal {P_R}$  is required to reach  the order of $\mathcal{O}(10^{-2})$.  Since the CMB observations have put  stringent constraints on  $\mathcal {P_R}$ only at the CMB scales, we can realize the production of abundant PBHs by enhancing the amplitude of the power spectrum of the curvature perturbations  about seven orders at small scales.
As $\mathcal {P_R} \propto 1/\epsilon$ with $\epsilon$ being the slow roll parameter, a natural way to amplify the curvature perturbations is to include an ultra-slow-roll period during inflation. Flattening the inflationary potential can reduce the rolling speed of the inflaton, which gives rises to an ultra-slow-roll inflation~\cite{Germani2017,Motohashi2017,Ezquiaga2017,H. Di2018,Ballesteros2018,Dalianis2019,Gao2018,Drees2021,C.Fu2020, Xu2020, Lin2020, Dalianis2021, Yi2021,Gao2021, Yi2021b, TGao2021, Solbi2021, Gao2021b, Solbi2021b, Zheng2021,Teimoori2021a, Cai2021, Wang2021,Karam2022}.   The ultra-slow-roll inflation can also be achieved  via slowing down the inflaton by gravitationally enhancing  friction~\cite{Fuchengjie2019,fuchengjie2020,Dalianis(2020),Teimoori2021, Heydari2022, Heydari2022b}.  Moreover,  some other mechanisms, such as parametric resonance~\cite{yfcai2018,yfcai2019,c.chen2019,c.chen2020,Addazi2022,Cai2020},  have also been proposed to amplify the curvature perturbations. 

When the amplified curvature perturbations  reenter the Hubble horizon during the radiation- or matter-dominated era,  they will not only generate the PBHs,  but also lead simultaneously to large scalar  metric perturbations, which  become an effective source of background gravitational waves. These gravitational waves, called the scalar induced gravitational waves (SIGWs), may be detectable by the future GW projects such as LISA~\cite{lisa}, Taiji~\cite{taiji}, TianQin~\cite{tianqin} and PTA~\cite{pta1,pta2,pta3,pta4}. 

When we assess the abundance of PBHs and the energy density of SIGWs, the curvature perturbations are assumed usually to be of a  Gaussian distribution. This is because 
the curvature perturbations generated during the standard slow-roll inflation are nearly Gaussian with negligible non-Gaussianity.    However, once the inflation departs from the slow-roll inflation or it is driven by the noncanonical fields, the primordial non-Gaussianity of the curvature perturbations may no longer be ignored.     The primordial non-Gaussianity  in the ultra-slow-roll inflation has been studied widely~\cite{Cai2019, QingGuoHuang2013,Zhang2021,F.Zhang2022,fengge2020,Chul-Moon Yoo2019, G.Franciolini2018,Matthew2022,  SamuelPassaglia2019,BravoRafael2018,Cai2018,VicenteAtal2018,VicenteAtal2019}, becuase the abundance of PBHs is extremely sensitive to the primordial  non-Gaussianity of  the curvature perturbations.   
For the PBHs generated from  inflation with  gravitationally enhanced friction  mechanism~\cite{Fuchengjie2019,Germani2011_1, Germani2011_2}, the primordial non-Gaussianity might be non-negligible too since the inflation field couples derivatively with the gravity and the rolling of the inflaton is ultra slow. In this paper we study,  in the ultra-slow-roll inflation achieved through  gravitationally enhanced friction, the non-Gaussianity of the curvature perturbations and its effect on the PBH abundance and the energy density of SIGWs.

The paper is organized   as follows: In Sec.~\ref{sec2}, we briefly review the inflation model with the nonminimal derivative coupling between inflation field and gravity.  Sec.~\ref{sec3} studies the primordial  non-Gaussianity of  the curvature perturbations.  In Sec.~\ref{sec4},   the effect of the non-Gaussianity  of  the curvature perturbations on the abundance of PBHs  and the energy density of SIGWs are assessed. Finally, we give our conclusions in Sec.~\ref{conclusion}.

\section{inflation with the gravitationally enhanced friction}
\label{sec2}
To enhance the friction term in the equation of motion of  the inflaton  through the gravity, we consider a nonminimal derivative coupling between the inflaton field $\phi$ and gravity,  with the action given by
 \begin{align}\label{action}
	\mathcal{S}=\int d^{4} x \sqrt{-g}\left[\frac{M_{\mathrm{pl}}^2}{2} R-\frac{1}{2}\left(g^{\mu \nu}-\frac{1}{M_{\mathrm{pl}}^2}  \theta(\phi) G^{\mu \nu}\right) \nabla_{\mu} \phi \nabla_{\nu} \phi-V(\phi)\right],
 \end{align}
where  $  M_{\mathrm{pl}} $ is the reduced Planck mass, and $g$ is the determinant of the metric tensor $g_{\mu\nu}$,  $R$ is the Ricci scalar, $G_{\mu\nu}$ is the Einstein tensor,  $\theta(\phi)$  is the coupling function, and $V(\phi)$ is the potential of the scalar inflaton field.

In the spatially flat Friedmann-Robertson-Walker background
\begin{align}\label{Jlagrange}
d s^{2}=-d t^{2}+a(t)^{2} d \mathbf{x}^{2}
\end{align}
with $a(t)$ being the scale factor, one can obtain, from the action (\ref{action}),  the background equations
\begin{align}\label{Jlagrange} 
	 3 H^{2}=\frac{1}{M_{\mathrm{pl}}^{2}}\left[\frac{1}{2}\left(1+ \frac{9}{M_{\mathrm{pl}}^{2}} \theta(\phi) H^{2}  \right) \dot{\phi}^{2}+V(\phi)\right],
\end{align}
\begin{align}\label{Jlagrange} 
	-2 \dot{H}=\frac{1}{M_{\mathrm{pl}}^{2}}\left[\left(1+\frac{3}{M_{\mathrm{pl}}^{2}} \theta(\phi) H^{2}-\frac{1}{M_{\mathrm{pl}}^{2}} \theta(\phi) \dot{H}\right) \dot{\phi}^{2}- \frac{1}{M_{\mathrm{pl}}^{2}} \theta_{, \phi} H \dot{\phi}^{3} - \frac{2}{M_{\mathrm{pl}}^{2}}  \theta(\phi) H \dot{\phi} \ddot{\phi}\right],
\end{align}
\begin{align}\label{Jlagrange} 
	\left(1+ \frac{3}{M_{\mathrm{pl}}^{2}} \theta(\phi) H^{2}\right) \ddot{\phi}+\left[1+\frac{1}{M_{\mathrm{pl}}^{2}} \theta(\phi)\left(2 \dot{H}+3 H^{2}\right)\right] 3 H \dot{\phi} +  \frac{3}{2M_{\mathrm{pl}}^{2}} \theta_{, \phi} H^{2} \dot{\phi}^{2}+V_{, \phi}=0 \, ,
\end{align}
where an overdot denotes the derivative with respective to the cosmic time $t$, $H=\frac{\dot{a}}{a}$ is the Hubble parameter,  $\theta_{, \phi}= d\theta/d\phi$, and $V_{, \phi}= dV/d\phi$. 

To describe the slow-roll inflation, we define the slow-roll parameters  
\begin{align}
\epsilon &=-\frac{\dot{H}}{H^{2}}, \quad \delta_{\phi} =\frac{\ddot{\phi}}{H \dot{\phi}}, \nonumber\\
\delta_{X} &=\frac{ \dot{\phi}^{2}}{2  {M_{\mathrm{pl}}^{2}}H^{2}}, \quad \delta_{D} =\frac{  \theta \dot{\phi}^{2}}{4{M_{\mathrm{pl}}^{4}}}.
\end{align}
When $\{\epsilon, |\delta_{\phi}|, \delta_{X}, \delta_{D} \}\ll 1$ are satisfied, the slow-roll inflation is obtained. 

\begin{table}[htbp]
  \caption{Three different  parameter sets for generating the $\mathcal{O}(10)M_\odot$, $\mathcal{O}(10^{-5})M_\odot$ and $\mathcal{O}(10^{-12})M_\odot$ PBHs, respectively.  $\phi_{*}$ and $N_*$ are the value of inflation field and e-folding number when the pivot scale $k_{*}=0.05\mathrm{Mpc}^{-1}$ exits the Hubble horizon. }\label{table1}
 \begin{tabular}{ccccccccccc}
  \hline
  \hline
  
  \hline
   &$\phi_{*}/M_{\mathrm{pl}}$ & $\phi_{c}/M_{\mathrm{pl}}$ & $ \sigma_{s} $ & $\lambda$ & $w$ &m  & $n_{s}$ & $r$ & $N_{*}$ & \\ 
  \hline
  Case 1 &$4.29$ & $4.02$ & $1.8 \times 10^{-9}$ & $6.68 \times 10^{-10}$& $ 3.70 \times 10^{16}$ & $6\times10^{8}$&0.971 &  0.0350   &64 & \\
  \hline
  Case 2 &$4.30$ & $3.63$ & $1.8 \times 10^{-9}$ & $6.60 \times 10^{-10}$& $ 4.08 \times 10^{16}$ &$8\times10^{8}$& 0.972 &  0.0340 &66& \\

  \hline
  Case 3 &$3.95$ & $2.95$ & $2.0 \times 10^{-9}$ & $7.40 \times 10^{-10}$& $ 5.19 \times 10^{16}$ & $9.5\times10^{8}$&0.971 &  0.0357 &68.5& \\

  \hline
  \hline
 \end{tabular}\\

\end{table}

In order to find the power spectrum of   the curvature perturbations, we need to derive the quadratic action for the curvature perturbations $\mathcal{R}$ from the action given in Eq.~(\ref{action}), which takes the form~\cite{A.D.Felice2011, S.Tsujikawa2012,Kobayashi2011} 
\begin{align}\label{S2}
S^{(2)}=\int d t d^{3} x a^{3} Q\left[\dot{\mathcal{R}}^{2}-\frac{c_{s}^{2}}{a^{2}}(\partial \mathcal{R})^{2}\right],
\end{align}
where
\bea\label{Q}
Q=\frac{w_{1}\left(4 w_{1} w_{3}+9 w_{2}^{2}\right)}{3 w_{2}^{2}},
\eea
\bea
c_{s}^{2}=\frac{3\left(2 w_{1}^{2} w_{2} H-w_{2}^{2} w_{4}+4 w_{1} \dot{w}_{1} w_{2}-2 w_{1}^{2} \dot{w}_{2}\right)}{w_{1}\left(4 w_{1} w_{3}+9 w_{2}^{2}\right)},
\eea
and 
\begin{align} \label{wi}
w_{1}&=M_{\mathrm{pl}}^{2}\left(1-2 \delta_{D}\right) , \nonumber \\
w_{2}&=2 H M_{\mathrm{pl}}^{2}\left(1-6 \delta_{D}\right), \nonumber \\
w_{3}&=-3 H^{2} M_{\mathrm{pl}}^{2}\left(3-\delta_{X}-36 \delta_{D}\right), \nonumber \\
w_{4}&=M_{\mathrm{pl}}^{2}\left(1+2 \delta_{D}\right).
\end{align}
From Eq.~(\ref{S2}), we obtain the Mukhanov-Sasaki equation
\begin{align}\label{action2}
u_{k}^{\prime \prime}+\left(c_{s}^{2} k^{2}-\frac{z^{\prime \prime}}{z}\right) u_{k}=0,
\end{align}
where $z^2 = 2a^2Q$, and $u_{k} = z\mathcal{R}_{k}$. Solving this  Mukhanov-Sasaki equation yields the power spectrum of the curvature perturbations
\begin{align}\label{power}
\mathcal{P}_\mathcal{R} \simeq \mathcal{P}_{\mathcal{R}{_0}} \left(1+\theta(\phi)\frac{V}{M_\mathrm{pl}^4}\right)
\end{align}
at the time when the comoving wave number exits the horizon, where  $ \mathcal{P}_{\mathcal{R}_0}= \frac{V^3}{12\pi^2M_\mathrm{pl}^6V_{,\phi}^2}$ is the power spectrum of the curvature perturbations in the  minimal coupling case. The scalar spectral index and the tensor-to-scalar ratio are given, respectively, by \cite{S.Tsujikawa2012}
\begin{align}\label{ns}
n_s\simeq1-\frac{1}{\mathcal{A}}\left[2\epsilon_V\left(4-\frac{1}{\mathcal{A}}\right)-2\eta_V\right]\;,
\end{align}
\begin{align}\label{r}
r\simeq\frac{16\epsilon_V}{\mathcal{A}}\;,
\end{align}
where $\epsilon_V=\frac{M_\mathrm{pl}^2}{2}\big(\frac{V_{,\phi}}{V} \big)^2$,  $\eta_V={M_\mathrm{pl}^2} \frac{V_{,\phi\phi}}{V}$ and $\mathcal{A}=1+\frac{3}{M_\mathrm{pl}^2}\theta(\phi) H^2$. 

 For the potential of the inflaton field, we choose the simple monomial  potential
\begin{align}\label{Potential}
 V(\phi)=\lambda M_{\mathrm{pl}}^{4-p}|\phi|^{p},
\end{align}
where $\lambda$ is a free parameter and the fractional power $p$ is set to be $p=2/5$ \cite{Silverstein2008}. 
 To amplify the curvature perturbations at the small scales to generate a sizable amount of  PBHs and at the same time to satisfy the strong constraint  on the tensor-to-scalar ratio ($r<0.036$) given by the BICEP/Keck collaboration~\cite{Ade2021},  the coupling function $\theta(\phi)$ is assumed to take the following form~\cite{Fuchengjie2019}
\begin{align}
\theta(\phi)=m+\frac{\omega}{\sqrt{\kappa^{2}\left(\frac{\phi-\phi_{c}}{\sigma_{s}}\right)^{2}+1}},
\end{align}
where $m$ is a coupling constant, which is introduced  to reduce  the tensor-to-scalar ratio  so as to  be consistent with the  BICEP/Keck CMB observations, $\omega$ and $\phi_{c}$ correspond to the peak height and position of the power spectrum of the curvature perturbations, and $\sigma_{s}$ describes the smoothing scale around $\phi=\phi_{c}$.

At the beginning of inflation,  the effect of the non-minimal derivative coupling can be neglected since $\phi$ deviates greatly from  $\phi_c$, and thus the inflationary prediction corresponds to that of the standard single-field slow-roll inflation with the  simple monomial  potential.   The friction will play a more and more important role with the inflaton field rolling toward $\phi_c$. The large friction reduces the rolling speed of the inflaton and leads to a period of ultra-slow-roll inflation.    Since the second term in parentheses of the r.h.s of Eq.~(\ref{power}) will become dominant,  the power spectrum will be enhanced.  The amplitude of the power spectrum of the curvature perturbations can be amplified  to be the order of $\mathcal{O}\left({10^{-2}}\right)$ during the ultra-slow-roll inflation. When these enhanced curvature perturbations reenter  the horizon during radiation- or matter-dominated era, a sizable amount of PBHs will be generated.

In order to use the PBHs to explain the binary black hole events detected by the LIGO/Virgo collaboration and  the ultrashort-timescale microlensing events in the OGLE data, and to make up all dark matter, we focus on the PBHs with mass around $\mathcal{O}(10)M_\odot$, $\mathcal{O}(10^{-5})M_\odot$, and $\mathcal{O}(10^{-12})M_\odot$,  and consider three  different parameter sets,  which are shown in Tab.~\ref{table1}. From this table, one can see that at the CMB scale the inflationary predictions are compatible with the BICEP/Keck CMB observations~\cite{Ade2021}. And the amplitude of the power spectrum of the curvature perturbations can be enhanced to be the $\mathcal{O}\left({10^{-2}}\right)$ order at the small scale to generate the  abundant PBHs, as shown in Tab.~\ref{table2}.

\section{primordial non-gaussianity}
\label{sec3}

 To study the primordial non-Gaussianity of  the curvature perturbations,  we need to calculate  the value of the bispectrum $B_{\mathcal{R}}$, which is related to the  three-point correlation function of  the curvature perturbations~\cite{C.T.Byrnes2010,P.Ade2016}
\begin{align}
\left\langle\hat{\mathcal{R}}_{\boldsymbol{k}_{1}} \hat{\mathcal{R}}_{\boldsymbol{k}_{2}} \hat{\mathcal{R}}_{\boldsymbol{k}_{3}}\right\rangle=(2 \pi)^{3} \delta^{3}\left(\boldsymbol{k}_{1}+\boldsymbol{k}_{2}+\boldsymbol{k}_{3}\right) B_{\mathcal{R}}\left(k_{1}, k_{2}, k_{3}\right)\;.
\end{align}
Using the in-in formula, we can calculate this three-point correlation function and   obtain the expression of the bispectrum $B_{\mathcal{R}} (k_1,k_2,k_3)$~\cite{Maldacena2003,FredericoArroja2011,X.Chen2007,DavidSeery2005,F.Zhang2022}
\begin{align}\label{bi}
B_{\mathcal{R}}\left(k_{1}, k_{2}, k_{3}\right)=  \Im \left[\mathcal{R}_{k_{1}}\left(t_{e}\right) \mathcal{R}_{k_{2}}\left(t_{e}\right) \mathcal{R}_{k_{3}}\left(t_{e}\right) \sum_{i=1}^{10} \mathcal{B}_{\mathcal{R}}^{i}\left(k_{1}, k_{2}, k_{3}\right)\right] \; .
\end{align}
Here $\Im$ represents  taking the  imaginary part,  $t_{e}$ denotes the time of the end of inflation, and the expressions of $\mathcal{B}_{\mathcal{R}}^{i}\left(k_{1}, k_{2}, k_{3}\right)$ are given in the appendix \ref{appa}. Then, we can derive the non-Gaussianity parameter $f_{\mathrm{NL}}$~\cite{C.T.Byrnes2010,Creminelli2002} 
\begin{align}
f_{\mathrm{NL}}\left(k_{1}, k_{2}, k_{3}\right)=\frac{5}{6} \frac{B_{\mathcal{R}}\left(k_{1}, k_{2}, k_{3}\right)}{P_{\mathcal{R}}\left(k_{1}\right) P_{\mathcal{R}}\left(k_{2}\right)+P_{\mathcal{R}}\left(k_{2}\right) P_{\mathcal{R}}\left(k_{3}\right)+P_{\mathcal{R}}\left(k_{3}\right) P_{\mathcal{R}}\left(k_{1}\right)}\; ,
\end{align}
where $P_{\mathcal{R}}(k)=\frac{2 \pi^{2}}{k^{3}} \mathcal{P}_{\mathcal{R}}(k)$.

We use the numerical method to calculate the value of $B_{\mathcal{R}} (k_1, k_2, k_3)$. Since the curvature perturbation $\mathcal{R}$ oscillates  rapidly  when it is in the horizon, a cutoff $\mathrm{e}^{\lambda k_{m}\left(\tau-\tau_{0}\right)}$ is introduced to reduce the error in numerical calculations \cite{X.Chen2007,X.Chen2008}, where $k_{m}$ is the largest value of $(k_1, k_2, k_3)$, $\lambda$ determines how much the integral will be suppressed, and $\tau_{0}$ is about several e-folding time before  the $k_{m}$ mode crosses the Hubble horizon. As the non-Gaussianity satisfies, in the squeezed limit, the consistency relation~\cite{Maldacena2003}
\begin{align}
\lim _{k_{3} \rightarrow 0} f_{\mathrm{NL}}\left(k_{1}, k_{2}, k_{3}\right)=\frac{5}{12} (1-n_{\mathrm{s}}) \quad \text { for } k_{1}=k_{2}\gg k_{3}\; ,
\end{align}
it can be used to verify the accuracy of the numerical calculation. 

\begin{figure*}
	\centering
		\includegraphics[width=0.52\linewidth]{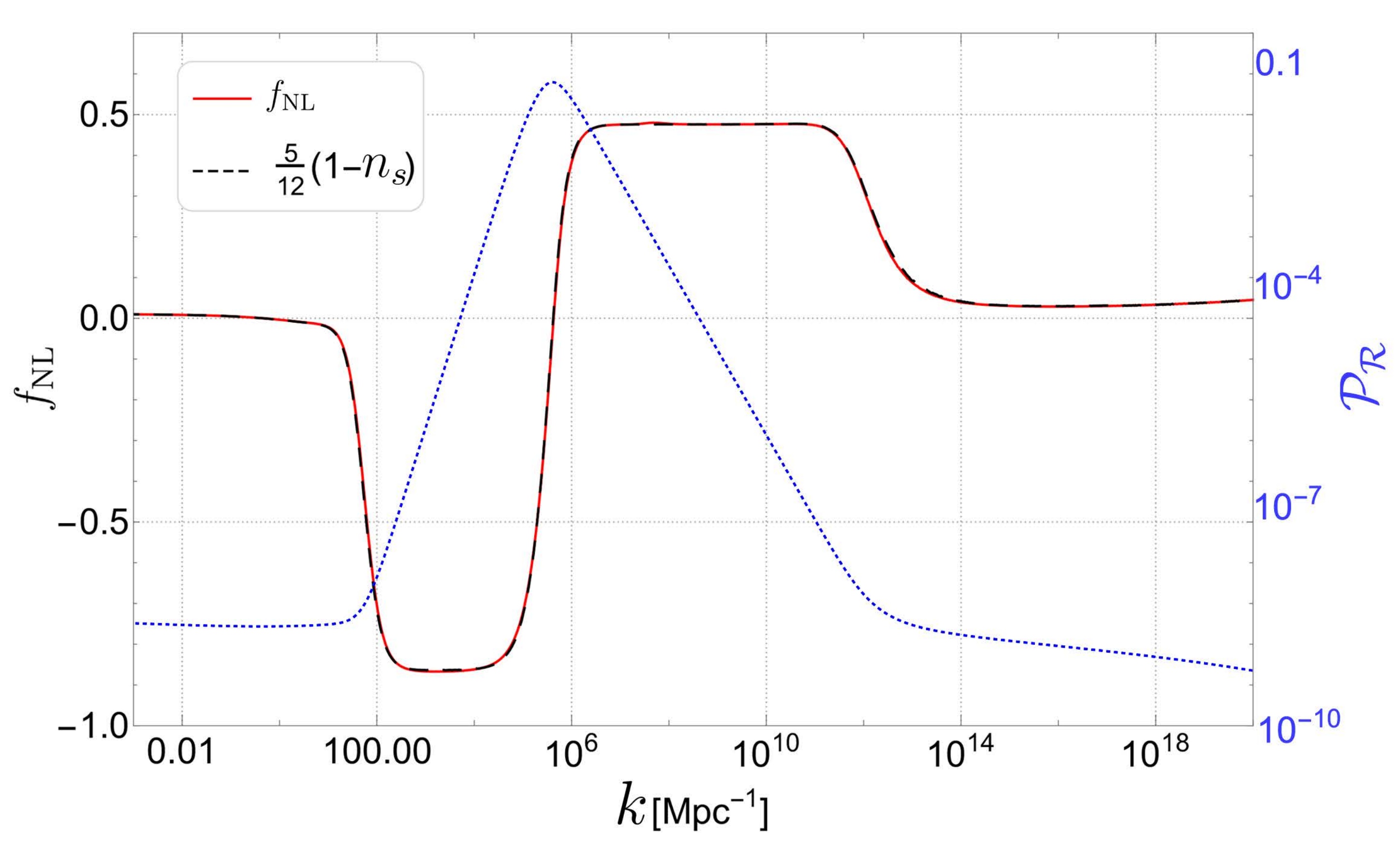}
	\caption{\label{Fig1}  $f_{\mathrm{NL}}$ in the squeezed limit with $k=k_{1}=k_{2}=10^{6}k_{3}$ for the case 1. Dashed line shows $\frac{5}{12}(1-n_{s})$ and dotted line represents the power spectrum of curvature perturbations.  }
\end{figure*}

\begin{figure*}
	\centering
		\includegraphics[width=0.52\linewidth]{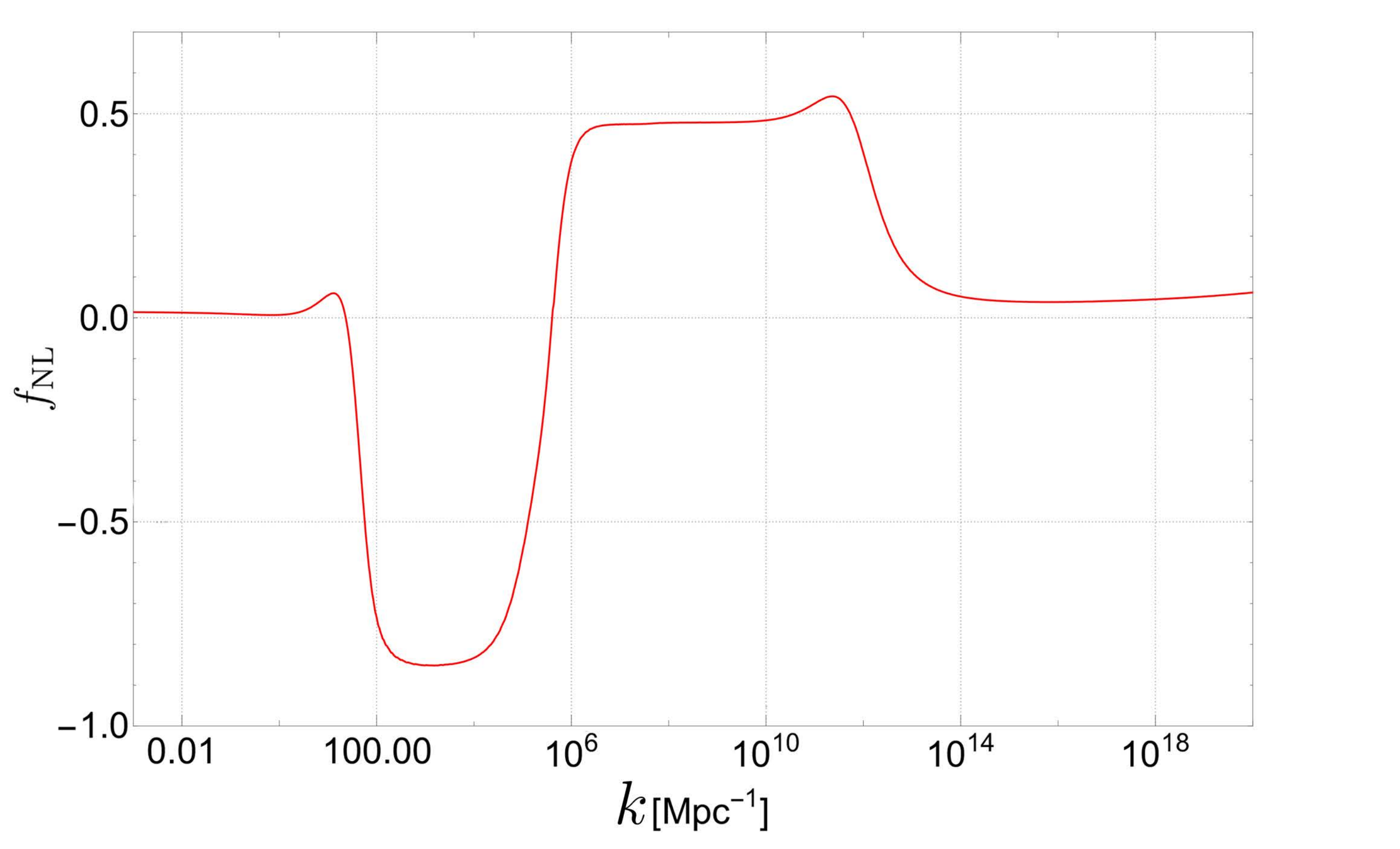}
		\caption{\label{Fig2}  $f_{\mathrm{NL}}$ in the equilateral limit ($k=k_1=k_2=k_3$) for the case 1.  }
\end{figure*}

Fig.~\ref{Fig1} shows our numerical results  in the squeezed limit for the case 1. The solid, dashed and dotted  lines represent  $f_\mathrm{NL}$, $\frac{5}{12} (1-n_{\mathrm{s}}) $ and the power spectrum, respectively.  One can see clearly   that $f_{\mathrm{NL}}$ satisfies  the non-Gaussianity consistency relation, which  demonstrates fully that our numerical calculation is very reliable. However, we must point out  here that this consistency relation could be violated if the ultra-slow-roll phase results  from a flattened potential~\cite{Cai2018,VicenteAtal2018, VicenteAtal2019}.
From Fig.~\ref{Fig1}, we find that   at the large scales,  the power spectrum of the curvature perturbations is nearly scale invariant,  the value of $f_{\mathrm{NL}}$ is around zero, and thus  non-Gaussianity is negligible, which is the prediction of the standard slow-roll inflation. With the decrease of scale, the power spectrum becomes to grow due to the slowing down of the inflaton as a result of  the gravitationally enhanced  friction, and accordingly the value of  $f_{\mathrm{NL}}$ drops sharply and then stabilizes at about $-0.86$. After several e-folding number, $f_\mathrm{NL}$ begins to increase rapidly and reaches near zero when the power spectrum reaches its maximum value at $k=k_\mathrm{peak}$.  Then, although the power spectrum decreases with the increase of $k$, $f_\mathrm{NL}$ will reach its maximum value, which is about $0.47$. Finally, with the ending of the ultra-slow-roll inflation, $f_\mathrm{NL}$ returns to about zero. 
 In figure~\ref{Fig2}, we plot the $f_\mathrm{NL}$ in the case of  the equilateral limit ($k_1=k_2=k_3$), and find that it has  features similar to that of  the squeezed limit case.

\section{Effect of non-Guassianity on PBHs and SIGWS}
\label{sec4}
\subsection{Non-Gaussian correction to the PBH abundance}

When the large enough curvature perturbations reenter the Hubble horizon during the radiation-dominated era, the gravity of  overdense regions can overcome the radiation pressure and thus these regions  will  collapse to form PBHs soon after their horizon entry.   The PBH mass relates with the horizon mass at the horizon entry of perturbations with the wave number $k$: 
\begin{align}
 M(k)=\gamma\frac{4\pi M_\mathrm{pl}^2}{H} \simeq M_{\odot}\left(\frac{\gamma}{0.2}\right)\left(\frac{g_{*}}{10.75}\right)^{-\frac{1}{6}}\left(\frac{k}{1.9 \times 10^{6}~ \mathrm{Mpc}^{-1}}\right)^{-2} \; ,
\end{align} 
where $\gamma$ denotes the ratio of the PBH mass  to the
horizon mass and indicates the efficiency of collapse, which is set to be $ \gamma \simeq(1 / \sqrt{3})^{3}$ \cite{Carr1975} in our analysis, and $g_{*}$ is the effective degrees of freedom in the energy densities at the PBH formation. We adopt $g_{*} = 106.75$ since the PBHs are assumed to form deep in the radiation-dominated.

 Based on the Press-Schechter theory \cite{Tada2019,Young2014}, the production rate of PBHs with mass $M(k)$ has the form
\bea
\beta(M)=\int_{\delta_{c}} \frac{d \delta}{\sqrt{2 \pi \sigma^{2}(M)}} e^{-\frac{\delta^{2}}{2 \sigma^{2}(M)}}=\frac{1}{2} \operatorname{erfc}\left(\frac{\delta_{c}}{\sqrt{2 \sigma^{2}(M)}}\right), 
\eea after assuming  that the probability distribution function of perturbations is Gaussian, where $\mathrm{erfc}$ is the complementary error function. $\delta_c$ is the threshold of the density perturbations for the PBH formation, which is chosen to be $\delta _{c} \simeq 0.4$ \cite{Musco2013,Harada2013} in our calculation of PBHs abundance.  $\sigma^{2}(M)$  has the form 
\bea \sigma^{2}(M(k))=  \frac{16}{81} \int d \ln q \; W^{2}\left(q k^{-1}\right) \left(q k^{-1}\right)^{4} \mathcal{P}_{\mathcal{R}}(q) \; ,
\eea
  which represents the coarse-grained density contrast with the smoothing scale $k$. Here $W$ is the window function. The current fractional energy density of PBHs with mass $M$ in dark matter is 
\begin{align}
f(M)  \equiv \frac{1}{\Omega_{\mathrm{DM}}} \frac{d \Omega_{\mathrm{PBH}}}{d \ln M} \simeq \frac{\beta(M)}{1.84 \times 10^{-8}}\left(\frac{\gamma}{0.2}\right)^{\frac{3}{2}}\left(\frac{10.75}{g_{*}}\right)^{\frac{1}{4}}\left(\frac{0.12}{\Omega_{\mathrm{DM}} h^{2}}\right)\left(\frac{M}{M_{\odot}}\right)^{-\frac{1}{2}},
\end{align}
where $\Omega_{\mathrm{DM}}$ is the current density parameter of dark matter, which is given to be $\Omega_{\mathrm{DM}}h^{2} \simeq  0.12$ by the Planck 2018 observations~\cite{Aghanim2020}. For the Gaussian distribution of the curvature perturbations, we obtain PBHs with masses around $\mathcal{O}\left({10}\right) M_\odot$, $\mathcal{O}\left({10^{-6}}\right) M_\odot$, and $\mathcal{O}\left({10^{-13}}\right)M_\odot$, respectively and their corresponding abundances, which are shown   in Tab.~\ref{table2}.

When the effect of  non-Guassianity of the curvature perturbations on the PBH abundance is considered, the mass fraction $\beta$   is corrected to be~\cite{G.Franciolini2018,Riccardi2021}
\begin{align}
\beta=\mathrm{e}^{\Delta_{3}} \beta^{G}\; ,
\end{align}
where  $\Delta _{3}$ is the 3rd cumulant,   which  has the form
\begin{align}\label{3rd}
\Delta_{3}=\frac{1}{3 !}\left(\frac{\delta_{c}}{\sigma}\right)^{2} S_{3} \delta_{c},
\end{align}
with $S_{3}$ being
\begin{align}
S_{3}=\frac{\left\langle\delta_{R}(\boldsymbol{x}) \delta_{R}(\boldsymbol{x}) \delta_{R}(\boldsymbol{x})\right\rangle}{\sigma^{4}}\;.
\end{align}
 For the Gaussian window function, $\left\langle\delta_{R}(\boldsymbol{x}) \delta_{R}(\boldsymbol{x}) \delta_{R}(\boldsymbol{x})\right\rangle$ can be obtained through calculating 
\bea
\left\langle\delta_{R}(\boldsymbol{x}) \delta_{R}(\boldsymbol{x}) \delta_{R}(\boldsymbol{x})\right\rangle&=&-64\left(\frac{4}{9}\right)^{3} \frac{2}{(2 \pi)^{4}} k^{6} \\ \nonumber 
& \times& \int_{0}^{\infty} d u \int_{0}^{\infty} d v \int_{|u-v|}^{u+v} d w u^{3} v^{3} w^{3} \mathrm{e}^{-u^{2}} \mathrm{e}^{-v^{2}} \mathrm{e}^{-w^{2}} B_{\mathcal{R}}(\sqrt{2} u k, \sqrt{2} v k, \sqrt{2} w k).
\eea
In \cite{Zhang2021}, it has been found that  $\Delta _{3}$ can be approximated as 
\begin{align}\label{3rd_a}
\Delta_{3} \approx \Delta_{3}^{a}(k_{\text{peak}}) = 23 \frac{\delta_{c}^{3}}{\mathcal{P}_{\mathcal{R}}\left(k_{\mathrm{peak}}\right)}~f_{\text {NL }}(k_{\mathrm{peak}}, k_{\mathrm{peak}}, k_{\mathrm{peak}} ).
\end{align}

  We numerically compute $\Delta _{3}$ and its approximation  $\Delta _{3}^{a}$. The results are shown in Tab.~\ref{table2} for three different cases. It is easy to see that $\Delta _{3}$ is of the order $\mathcal{O} (1)$, which is much larger than $\Delta _{3}^{a}$. Thus, the approximation given in  \cite{Zhang2021} is invalid for the model considered in the present paper.  We find that for all the cases the  non-Gaussianity promotes the formation of PBHs since their 3rd cumulants $\Delta _{3}$ are  positive.  The effect of non-Gaussianity on the PBH abundance is non-negligible since the value of $\beta$ is much larger  than $\beta^G$. 

\begin{table}[htbp]
  \caption{The numerical results for three cases given in Tab.~\ref{table1}. }\label{table2}
 \begin{tabular}{cccccccc}
  \hline
  \hline
   &$k_{\mathrm{peak}}/\mathrm{Mpc}^{-1}$ & ~~$\mathcal{P}_{\mathcal{R}}/10^{-2}$ & ~$f_{\mathrm{PBH}}^{G}$ &~~~$M_{\mathrm{PBH}}/M_{\odot}$ & $f_{\mathrm{NL}}$ & $\Delta _{3}^{a}$ &  $\Delta _{3}$\\
  \hline
  Case 1 &$4.46 \times 10^{5}$ & $4.91$ & $0.00455$ & $16.57$  & $~ ~0.0205$ & $~~0.619$ & $~4.218$ \\
  \hline
  Case 2 &$4.31 \times 10^{8}$ & $3.87$ & $0.0178$ &$~~1.91 \times 10^{-5}$ & $~~0.0234$ & $~~0.890$ & $~4.965$\\
  \hline
  Case 3 &$1.77 \times 10^{12}$ & $3.18$ & $0.924$ &~~~$1.12 \times 10^{-12} $ & $-0.0052$ & $-0.240$ & $~6.297$ \\
  \hline
  \hline
 \end{tabular}\\
\end{table}

\subsection{Non-Gaussian correction to energy density of SIGWs}

Associated with the PHB formation, the enhanced curvature perturbations  will lead to the large scalar metric perturbations, which  become the significant GW source  and emit abundant SIGWs. The current energy spectra of SIGWs can be expressed as~\cite{Kohri2018, Inomata2019}
\begin{align}
\Omega_{\mathrm{GW}, 0} h^{2}=0.83\left(\frac{g_{*}}{10.75}\right)^{-1 / 3} \Omega_{\mathrm{r}, 0} h^{2} \Omega_{\mathrm{GW}}\left(\tau_{c}, k\right),
\end{align}
where $\Omega_{\mathrm{r}, 0} h^{2}\simeq 4.2 \times  10^5$ is the current density parameter of radiation, and  
\begin{align}
\Omega_{\mathrm{GW}}\left(\tau_{c}, k\right)=& \frac{1}{12} \int_{0}^{\infty} d v \int_{|1-v|}^{|1+v|} d u\left(\frac{4 v^{2}-\left(1+v^{2}-u^{2}\right)^{2}}{4 u v}\right)^{2} \mathcal{P}_{\mathcal{R}}(k u) \mathcal{P}_{\mathcal{R}}(k v) \nonumber\\
&\times \left(\frac{3}{4 u^{3} v^{3}}\right)^{2}\left(u^{2}+v^{2}-3\right)^{2} \nonumber\\
&\times \bigg \{\left[-4 u v+\left(u^{2}+v^{2}-3\right) \ln \left|\frac{3-(u+v)^{2}}{3-(u-v)^{2}}\right|\right]^{2} \nonumber
\\ &+\pi^{2}\left(u^{2}+v^{2}-3\right)^{2} \Theta(v+u-\sqrt{3})\bigg\}\, ,
\end{align}
where $\tau_c$ represents the time when $\Omega_\mathrm{GW}$ stops to grow and $\Theta$ is the Heaviside theta function.  

When the   non-Gaussianity of the curvature perturbations is considered, $\mathcal{R}(\boldsymbol{x})$ has the expression~\cite{Verde2000,Komatsu2001}
\begin{align}
\mathcal{R}(\boldsymbol{x})=\mathcal{R}^{G}(\boldsymbol{x})+\frac{3}{5} f_{\mathrm{NL}}\left(\mathcal{R}^{G}(\boldsymbol{x})^{2}-\left\langle\mathcal{R}^{G}(\boldsymbol{x})^{2}\right\rangle\right)\, .
\end{align}
Clearly the curvature perturbation $\mathcal{R}$ consists of a Gaussian part $\mathcal{R}^G$ and a non-Gaussian one.  When this non-Gaussian correction is included, the power spectrum of the curvature perturbations should be modified to be   
\begin{align}
\mathcal{P}_{\mathcal{R}}(k)=\mathcal{P}_{\mathcal{R}}^{G}(k)+\mathcal{P}_{\mathcal{R}}^{N G}(k),
\end{align}
where 
\begin{align}
\mathcal{P}_{\mathcal{R}}^{N G}(k)=\left(\frac{3}{5}\right)^{2} \frac{k^{3}}{2 \pi} f_{\mathrm{NL}}^{2} \int d^{3} \boldsymbol{p} \frac{\mathcal{P}_{\mathcal{R}}^{G}(p)}{p^{3}} \frac{\mathcal{P}_{\mathcal{R}}^{G}(|\boldsymbol{k}-\boldsymbol{p}|)}{|\boldsymbol{k}-\boldsymbol{p}|^{3}}.
\end{align}
For the model considered in this paper, the result in the preceding  subsection has shown that  the absolute value of $f_{\mathrm{NL}}$ is less than one and it is near zero when $k=k_\mathrm{peak}$. Furthermore the maximum value of $\mathcal{P}_{\mathcal{R}}^{G}$ has the order of $\mathcal{O}(0.01)$. Thus, we can  assess easily that $\mathcal{P}_{\mathcal{R}}^{N G}(k)$ is much less than $\mathcal{P}_{\mathcal{R}}^{G}(k)$ since the order of its maximum should be less than $\mathcal{O}(10^{-4})$, which indicates that   the contribution of non-Gaussianity of the curvature perturbations on the energy density of SIGWs is negligible.  

 \section{conclusions}
 \label{conclusion}
 To generate a sizable amount of PBHs requires that the amplitude of the power spectrum of the curvature perturbations is enhanced to reach the $\mathcal{O}(0.01)$ order. A simple way to enhance the curvature perturbations during inflation is to reduce the rolling speed of inflaton to achieve an ultra-slow-roll inflation, which can be realized by flattening the inflationary potential or increasing the gravitational friction. Since the ultra-slow-roll inflation deviates apparently from the standard slow-roll inflation, the non-Gaussianity of the curvature perturbations might be very large and has significant effects on the abundance of PBHs and the energy density of SIGWs  although it is negligible in the standard slow-roll inflation.
 
  In this paper we study the non-Gaussianity of the curvature perturbations in the ultra-slow-roll inflation resulting from gravitationally enhanced friction. We find that at the large scales  where the power spectrum of the curvature perturbations is nearly scale invariant,  the non-Gaussianity is negligible. The power spectrum  grows with the decrease of scale due to that the friction slows down the inflaton, and correspondingly the value of the non-Gaussianity parameter $f_{\mathrm{NL}}$ drops sharply and then stabilizes at a value in several e-folding number. Before the power spectrum reaches its peak,   $f_\mathrm{NL}$ begins to increase rapidly.   When the power spectrum is at  its maximum value, we find that $f_\mathrm{NL}$  is nearly  zero.  Then, $f_\mathrm{NL}$  will reach its maximum value with the increase of wave number $k$.  Finally, with the ending of the ultra-slow-roll inflation, $f_\mathrm{NL}$ returns to about zero.   For three different cases, which correspond to that the PBHs can be used to explain the LIGO/Virgo GW events and the six ultrashort-timescale microlensing events in the OGLE data, and make up all dark matter, respectively,   we obtain that the non-Gaussianity will promote  the generation of PBHs, while its influence  on SIWGs is   negligible.


\appendix
\section{The expressions of $\mathcal{B}_{\mathcal{R}}^{i}\left(k_{1}, k_{2}, k_{3}\right)$ in Eq.~(\ref{bi}) }\label{appa}
In order to compute bispectrum, we need derive the cubic action of the curvature perturbations from the action given in Eq.~(\ref{action}) \cite{A.D.Felice2011, Felice2011}
\bea
S_{3}&=& \int d t d^{3} x\left\{a^{3} \mathcal{C}_{1} \mathcal{R} \dot{\mathcal{R}}^{2}+a \mathcal{C}_{2} \mathcal{R}(\partial \mathcal{R})^{2}+a^{3} \mathcal{C}_{3} \dot{\mathcal{R}}^{3}+a^{3} \mathcal{C}_{4} \dot{\mathcal{R}}\left(\partial_{i} \mathcal{R}\right)\left(\partial_{i} \chi\right)\right.\nonumber\\
&&+a^{3} \mathcal{C}_{5} \partial^{2} \mathcal{R}(\partial \chi)^{2}+a \mathcal{C}_{6} \dot{\mathcal{R}}^{2} \partial^{2} \mathcal{R}+\left(\mathcal{C}_{7} / a\right)\left[\partial^{2} \mathcal{R}(\partial \mathcal{R})^{2}-\mathcal{R} \partial_{i} \partial_{j}\left(\partial_{i} \mathcal{R}\right)\left(\partial_{j} \mathcal{R}\right)\right] \nonumber\\
&&+a \mathcal{C}_{8}\left[\partial^{2} \mathcal{R} \partial_{i} \mathcal{R} \partial_{i} \chi-\mathcal{R} \partial_{i} \partial_{j}\left(\partial_{i} \mathcal{R}\right)\left(\partial_{j} \chi\right)\right]+\left.\mathcal{F}_{1} \frac{\delta \mathcal{L}_{2}}{\delta \mathcal{R}} \right\},
\eea
where
\bea
\mathcal{F}_{1}&=& \frac{A_{4}}{w_{1}^{2}}\left\{\left(\partial_{k} \mathcal{R}\right)\left(\partial_{k} \chi \right)-\partial^{-2} \partial_{i} \partial_{j}\left[\left(\partial_{i} \mathcal{R}\right)\left(\partial_{j} \chi \right)\right]\right\} \nonumber\\
&&+q_{1} \mathcal{R} \dot{\mathcal{R}} - \frac{A_4 }{ a^{2} w_{2} } \times\left\{(\partial \mathcal{R})^{2}-\partial^{-2} \partial_{i} \partial_{j}\left[\left(\partial_{i} \mathcal{R}\right)\left(\partial_{j} \mathcal{R}\right)\right]\right\} \nonumber ,
\eea
 $\partial^{2} \chi=Q \dot{\mathcal{R}}$,    $\mathcal{L}_{2}$ is quadratic Lagrangian given in Eq.~(\ref{S2}),  $w_i$ and $Q$ are given in Eqs.~(\ref{wi}) and (\ref{Q}), respectively,  and the dimensionless coefficients  $\mathcal{C}_{i}$ with $i = 1 - 8$ are 
\bea
\mathcal{C}_{1} &=&\frac{1}{M_{\mathrm{pl}}^{2}}\left[3Q+q_{1}(\dot{Q}+3 H Q)-Q \dot{q}_{1}\right], \\ 
\mathcal{C}_{2}&=&\frac{1}{M_{\mathrm{pl}}^{2}}\left[A_5 +\frac{1}{a} \frac{d}{d t}\left( \frac{2aQw_{1}}{w_{2}} \right)\right] ,\\
\mathcal{C}_{3} &=&\frac{1}{M_{\mathrm{pl}}}\left[A_{1}+A_{3} \frac{Q}{w_{1}}-q_{1} Q\right], \\
\mathcal{C}_{4} &=&\frac{Q}{w_{1}}\left[-\frac{1}{2} -w_{1} \frac{d}{d t}\left(\frac{A_4}{w_{1}^{2}}\right)+\frac{3 H A_4}{w_{1}}\right], \\
\mathcal{C}_{5} &=&\frac{M_{\mathrm{pl}}^{2}}{2}\left[\frac{3}{2 w_1}-\frac{d}{d t}\left(\frac{A_4}{w_{1}^{2}}\right)+\frac{3 H A_4}{w_{1}^{2}}\right] ,\\
\mathcal{C}_{6}&=&A_{2}- \frac{2w_{1}A_{3}}{w_{2}}, \\
 \mathcal{C}_{7}&=&q_{3}+\frac{2 A_4 Q c_{s}^{2}}{ w_{2}} , \\ 
\mathcal{C}_{8}&=&M_{\mathrm{pl}}\left[\frac{q_{2}}{2}-\frac{2 c_{s}^{2} A_4 Q}{w_{1}^{2}}\right].
\eea
Here
\bea
A_{1}&=&\frac{3w_{1} }{w^{3}_{2}}\bigg[ 8H^{2}\left (M_{\mathrm{pl}}^{2}-5\dot{\phi}^{2}\theta\left (\phi \right) \right )w_{1}^{2} +8H\left ( 3\dot{\phi }^{2}\theta \left( \phi \right) -M_{\mathrm{pl}}^{2}\right )w_{1}w_{2}   +\left (2M_{\mathrm{pl}}^{2}-3\dot{\phi }^{2}\theta \left (\phi \right ) \right )w_{2}^{2}\bigg ]  ,\nonumber \\
 A_{2} &=& - 4\dot{\phi }^{2}\theta(\phi) \frac{w_{1}^{2}   }{w^{2}_{2}} ,  \nonumber  \\
 A_{3} &=& \frac{2w_{1} }{w^{2}_{2}} \left [4Hw_{1}\left (M_{\mathrm{pl}}^{2}-3\dot{\phi }^{2}\theta \left ( \phi \right )\right )+w_2\left (3\dot{\phi }^{2}\theta \left ( \phi \right )-2M_{\mathrm{pl}}^{2}\right)\right ] , \nonumber  \\
A_{4} &=& \frac{w_{1} }{2w_{2}}  \left (3\dot{\phi }^{2}\theta \left ( \phi \right )-2 M_{\mathrm{pl}}^{2}\right ), \nonumber  \\  
A_{5}&=&\frac{2\dot{w}_{2}w_{1}^{2}+w_{2}\left (w_{2}w_{4}-4\dot{w}_{1}w_{1}-2Hw_{1}^{2}  \right ) }{w_{2}^{2}}  , \nonumber  
\eea
and
\bea
q_{1} &=&  -\frac{2w_{1}}{c_{s}^{2}w_{2}}  ,\nonumber \\
q_{2} &=&  a^{2}\frac{d}{dt} \left [ \frac{w_{1}\left (4M_{\mathrm{pl}}^{2}-6\dot{\phi }^{2}\theta \left (\phi \right ) \right )}{a^{2}w_{2}^{2}}   \right ]-\frac{4w_{1}}{w_{2}} , \nonumber \\ \nonumber
q_{3} &= & \frac{2w_{1}^{3}}{3w_{2}^{2}}-\frac{a}{3}\frac{d}{dt}  \left [\frac{2w_{1}^{3}\left (3\dot{\phi }^{2}\theta (\phi )-2M_{\mathrm{pl}}^{2}\right )}{aw_{2}^{3}}   \right] .
\eea

Using the in-in formula,   one can obtain the three-point correlation function  from  the cubic action of the curvature perturbations. The analytical expression is shown in  Eq.~(\ref{bi}), in which  $\mathcal{B}_{\mathcal{R}}^{i}\left(k_{1}, k_{2}, k_{3}\right)$  have the forms
\begin{align}
\mathcal{B}_{\mathcal{R}}^{1}\left(k_{1}, k_{2}, k_{3}\right)=-4 \int_{t_{i}}^{t_{e}} dt \; a^{3} \; \mathcal{C}_{1}\left(\mathcal{R}_{k_{1}}^{*}(t) \dot{\mathcal{R}}_{k_{2}}^{*}(t) \dot{\mathcal{R}}^{*}_{k_{3}}(t)+\text { perm }\right),
\end{align} 
\begin{align}
\mathcal{B}_{\mathcal{R}}^{2}\left(k_{1}, k_{2}, k_{3}\right)=4 \int_{t_{i}}^{t_{e}} d t \;a\; \mathcal{C}_{2}\left[\left(\boldsymbol{k}_{1} \cdot \boldsymbol{k}_{2}+\boldsymbol{k}_{1} \cdot \boldsymbol{k}_{3}+\boldsymbol{k}_{2} \cdot \boldsymbol{k}_{3}\right) \mathcal{R}_{k_{1}}^{*}(t) \mathcal{R}_{k_{2}}^{*}(t) \mathcal{R}_{k_{3}}^{*}(t)\right],
\end{align} 
\begin{align}
\mathcal{B}_{\mathcal{R}}^{3}\left(k_{1}, k_{2}, k_{3}\right)=-12 \int_{t_{i}}^{t_{e}} d t \; a^{3} \; \mathcal{C}_{3} \dot{\mathcal{R}}_{k_{1}}^{*}(t) \dot{\mathcal{R}}_{k_{2}}^{*}(t) \dot{\mathcal{R}}_{k_{3}}^{*}(t),
\end{align} 
\begin{align}
\mathcal{B}_{\mathcal{R}}^{4}\left(k_{1}, k_{2}, k_{3}\right)=-2 \int_{t_{i}}^{t_{e}} d t \; a^{3}\; \mathcal{C}_{4} \;Q\left[\left(\frac{\boldsymbol{k}_{1} \cdot \boldsymbol{k}_{2}}{k_{2}^{2}}+\frac{\boldsymbol{k}_{1} \cdot \boldsymbol{k}_{3}}{k_{3}^{2}}\right) \mathcal{R}_{k_{1}}^{*}(t) \dot{\mathcal{R}}_{k_{2}}^{*}(t) \dot{\mathcal{R}}_{k_{3}}^{*}(t)+\text { perm }\right],
\end{align} 
\begin{align}
\mathcal{B}_{\mathcal{R}}^{5}\left(k_{1}, k_{2}, k_{3}\right)=-4  \int_{t_{i}}^{t_{e}} d t \;a^{3} \; \mathcal{C}_{5} \; Q^{2}\left[\frac{k_{1}^{2} \boldsymbol{k}_{2} \cdot \boldsymbol{k}_{3}}{k_{2}^{2} k_{3}^{2}} \mathcal{R}_{k_{1}}^{*}(t) \dot{\mathcal{R}}_{k_{2}}^{*}(t) \dot{\mathcal{R}}_{k_{3}}^{*}(t)+\text { perm }\right],
\end{align} 
\begin{align}
\mathcal{B}_{\mathcal{R}}^{6}\left(k_{1}, k_{2}, k_{3}\right)=4  \int_{t_{i}}^{t_{e}} d t \;a\; \mathcal{C}_{6}\left[k_{1}^{2} \mathcal{R}_{k_{1}}^{*}(t) \dot{\mathcal{R}}^{*} k_{2}(t) \dot{\mathcal{R}}^{*} k_{3}(t)+\text { perm }\right],
\end{align} 
\begin{align}
\mathcal{B}_{\mathcal{R}}^{7}\left(k_{1}, k_{2}, k_{3}\right)=-4  \int_{t_{i}}^{t_{e}} d t \; \mathcal{C}_{7} / a\left[\left(k_{1}^{2} \boldsymbol{k}_{2} \cdot \boldsymbol{k}_{3}+k_{2}^{2} \boldsymbol{k}_{1} \cdot \boldsymbol{k}_{3}+k_{3}^{2} \boldsymbol{k}_{1} \cdot \boldsymbol{k}_{2}\right) \mathcal{R}_{k_{1}}^{*}(t) \mathcal{R}_{k_{2}}^{*}(t) \mathcal{R}_{k_{3}}^{*}(t)\right],
\end{align} 
\begin{align}
\mathcal{B}_{\mathcal{R}}^{8}\left(k_{1}, k_{2}, k_{3}\right)=2 \int_{t_{i}}^{t_{e}} d t\; \mathcal{C}_{7} / a & {\left[\left(k_{2}^{2} \boldsymbol{k}_{2} \cdot \boldsymbol{k}_{3}+k_{3}^{2} \boldsymbol{k}_{3} \cdot \boldsymbol{k}_{2}+k_{1}^{2} \boldsymbol{k}_{1} \cdot \boldsymbol{k}_{2}+k_{1}^{2} \boldsymbol{k}_{1} \cdot \boldsymbol{k}_{3}\right.\right.} \nonumber\\
&\left.\left.+k_{2}^{2} \boldsymbol{k}_{2} \cdot \boldsymbol{k}_{1}+k_{3}^{2} \boldsymbol{k}_{3} \cdot \boldsymbol{k}_{1}\right) \mathcal{R}_{k_{1}}^{*}(t) \mathcal{R}_{k_{2}}^{*}(t) \mathcal{R}_{k_{3}}^{*}(t)\right],
\end{align} 
\begin{align}
\mathcal{B}_{\mathcal{R}}^{9}\left(k_{1}, k_{2}, k_{3}\right)=2  \int_{t_{i}}^{t_{e}} d t \; a \; \mathcal{C}_{8} \; Q\left[\left(\frac{k_{1}^{2} \boldsymbol{k}_{2} \cdot \boldsymbol{k}_{3}}{k_{3}^{2}}+\frac{k_{2}^{2} \boldsymbol{k}_{1} \cdot \boldsymbol{k}_{3}}{k_{3}^{2}}\right) \mathcal{R}_{k_{1}}^{*}(t) \mathcal{R}_{k_{2}}^{*}(t) \dot{\mathcal{R}}_{k_{3}}^{*}(t)+\text { perm }\right],
\end{align} 
\begin{align}
\mathcal{B}_{\mathcal{R}}^{10}\left(k_{1}, k_{2}, k_{3}\right)=-2 \int_{t_{i}}^{t_{e}} d t \; a \; \mathcal{C}_{8} \; Q\left[\left(\frac{k_{2}^{2} \boldsymbol{k}_{2} \cdot \boldsymbol{k}_{3}}{k_{3}^{2}}+\frac{k_{1}^{2} \boldsymbol{k}_{1} \cdot \boldsymbol{k}_{3}}{k_{3}^{2}}\right) \mathcal{R}_{k_{1}}^{*}(t) \mathcal{R}_{k_{2}}^{*}(t) \dot{\mathcal{R}}_{k_{3}}^{*}(t)+\operatorname{perm}\right].
\end{align} 

\begin{acknowledgments}
We appreciate very much the insightful comments and helpful suggestions by the anonymous referee. L. Chen is grateful to Dr. Fengge Zhang for his valuable help in the non-Gaussian numerical computation.
This work is supported by the National Key Research and Development Program of China Grant No. 2020YFC2201502, and  by the National Natural Science Foundation of China under Grants No. 12275080, No. 12075084,   and No. 11805063. 

\end{acknowledgments}


\begin{thebibliography}{99}
 
\bibitem{Guth1980}
A. H. Guth,
\href{https://doi.org/10.1103/PhysRevD.23.347}
{Phys. Rev. D \textbf{23}, 347 (1981)}.

\bibitem{Linde1982}
A. D. Linde,
\href{https://doi.org/10.1016/0370-2693(82)91219-9}
{Phys. Lett. \textbf{108B}, 389 (1982)}.

\bibitem{Starobinsky1980}
A. A. Starobinsky,
\href{https://doi.org/10.1016/0370-2693(80)90670-X}
{Phys. Lett.  \textbf{91B}, 99 (1980)}.


\bibitem{Albrecht1982}
A. Albrecht and P. J. Steinhardt,
\href{https://doi.org/10.1103/PhysRevLett.48.1220}
{Phys. Rev. Lett. \textbf{48}, 1220 (1982)}.

\bibitem{Aghanim2020}
N. Aghanim et al. (Planck),
\href{https://doi.org/10.1051/0004-6361/201833910}
{Astron. Astrophys. \textbf{641}, A6 (2020)}.

\bibitem{Hawking1971}
S. Hawking,
\href{https://doi.org/10.1093/mnras/152.1.75}
{Mon. Not. R. Astron. Soc. \textbf{152}, 75 (1971)}. 

\bibitem{Carr1974}
B. J. Carr and S. W. Hawking, 
\href{https://doi.org/10.1093/mnras/168.2.399}
{Mon. Not. R. Astron. Soc. \textbf{168}, 399 (1974)}.

\bibitem{Carr1975} 
B. J. Carr,
\href{https://doi.org/10.1086/153853}
{ Astrophys. J. \textbf{201}, 1 (1975)}.

\bibitem{Khlopov2007} M. Y. Khlopov, 
\href{https://doi.org/10.1088/1674-4527/10/6/001}{Res. Astron. Astrophys. \textbf{10}, 495 (2010)}.

\bibitem{lg1}
S. Bird, I. Cholis, J. B. Mu\~{n}oz, Y. Ali-Ha\"{i}moud, M. Kamionkowski, E. D. Kovetz, A. Raccanelli, and A. G. Riess,
\href{https://doi.org/10.1103/PhysRevLett.116.201301}
{Phys. Rev. Lett. \textbf{116}, 201301 (2016)}.

\bibitem{lg2}
S. Clesse and J. Garc\'{i}a-Bellido,
\href{https://doi.org/10.1016/j.dark.2016.10.002}	
{Phys. Dark Univ. \textbf{15}, 142 (2017)}.

\bibitem{lg3}
M. Sasaki, T. Suyama, T. Tanaka, and S. Yokoyama,
\href{https://doi.org/10.1103/PhysRevLett.121.059901}
{Phys. Rev. Lett. \textbf{117}, 061101 (2016)}.
\bibitem{lg4}
B. Carr, F. Kuhnel, and M. Sandstad,
\href{https://doi.org/10.1103/PhysRevD.94.083504}
{Phys. Rev. D \textbf{94}, 083504 (2016)}.

\bibitem{P.Mroz2017}
P. Mr\'oz, A. Udalski,  J. Skowron, et al.,
\href{https://doi.org/10.1038/nature23276}
{Nature (London) \textbf{548}, 183 (2017)}.



\bibitem{H.Niikura2019}
H. Niikura, M. Takada, S. Yokoyama, T. Sumi, and S. Masaki,
\href{https://doi.org/10.1103/PhysRevD.99.083503}
{Phys. Rev. D \textbf{99}, 083503 (2019)}.




\bibitem{A.Katz2018}
A. Katz, J. Kopp, S. Sibiryakov, and W. Xue,
\href{https://doi.org/10.1088/1475-7516/2018/12/005}
{J. Cosmol. Astropart. Phys. \textbf{12} (2018) 005}.


\bibitem{A.Barnacka2012}
A. Barnacka, J. F. Glicenstein, and R. Moderski,
\href{https://doi.org/10.1103/PhysRevD.86.043001}
{Phys. Rev. D \textbf{86}, 043001 (2012)}.

\bibitem{P.W.Graham2015}
P. W. Graham, S. Rajendran, and J. Varela,
\href{https://doi.org/10.1103/PhysRevD.92.063007}
{Phys. Rev. D \textbf{92}, 063007 (2015)}.



\bibitem{H.Niikura20191}
H. Niikura, M. Takada, N. Yasuda,  et al.,
\href{https://doi.org/10.1038/s41550-019-0723-1}
{Nat. Astron. \textbf{3}, 524 (2019)}.

\bibitem{Belotsky2014}
K. M. Belotsky, A. D. Dmitriev, E. A. Esipova, V. A. Gani, A. V. Grobov, 
M. Y. Khlopov, A.A.Kirillov, S. G. Rubin, and I. V. Svadkovsky,
\href{https://doi.org/10.1142/S0217732314400057}{
Mod. Phys. Lett. A \textbf{ 29},   1440005 (2014)}.


\bibitem{Germani2017}
C. Germani and T. Prokopec,
\href{https://doi.org/10.1016/j.dark.2017.09.001}
{Phys. Dark Univ. \textbf{18}, 6 (2017)}.

\bibitem{Motohashi2017}
H. Motohashi and W. Hu,
\href{https://doi.org/10.1103/PhysRevD.96.063503}
{Phys. Rev. D \textbf{96}, 063503 (2017)}.

\bibitem{Ezquiaga2017}
J. M. Ezquiaga, J. Garcia-Bellido, and E. Ruiz Morales,
\href{https://doi.org/10.1016/j.physletb.2017.11.039}
{Phys. Lett. \textbf{039B}, 11 (2017)}.

\bibitem{H. Di2018}
H. Di and Y. Gong,
\href{https://doi.org/10.1088/1475-7516/2018/07/007}
{J. Cosmol. Astropart. Phys. \textbf{07} (2018) 007}.

\bibitem{Ballesteros2018}
G. Ballesteros and M. Taoso,
\href{https://doi.org/10.1103/PhysRevD.97.023501}
{Phys. Rev. D \textbf{97}, 023501 (2018)}.

\bibitem{Dalianis2019}
I. Dalianis, A. Kehagias, and G. Tringas,
\href{https://doi.org/10.1088/1475-7516/2019/01/037}
{J. Cosmol. Astropart. Phys. \textbf{01} (2019) 037}.

\bibitem{Gao2018}
T. J. Gao and Z. K. Guo,
\href{https://doi.org/10.1103/PhysRevD.98.063526}
{Phys. Rev. D \textbf{98}, 063526 (2018)}.

\bibitem{Drees2021}
M. Drees and Y. Xu,
\href{https://doi.org/10.1140/epjc/s10052-021-08976-2}
{Eur. Phys. J. C \textbf{81}, 182 (2021)}.

\bibitem{C.Fu2020}
C. Fu, P. Wu, and H. Yu,
\href{https://doi.org/10.1103/PhysRevD.102.043527}
{Phys. Rev. D \textbf{102}, 043527 (2020)}.

\bibitem{Xu2020}
W. Xu, J. Liu, T. Gao, and Z. Guo,
\href{https://doi.org/10.1103/PhysRevD.101.023505}
{Phys. Rev. D \textbf{101},  023505 (2020)}.


\bibitem{Lin2020}
J. Lin, Q. Gao, Y. Gong, Y. Lu, and C. Zhang,
\href{https://doi.org/10.1103/PhysRevD.101.103515}
{Phys. Rev. D \textbf{101}, 103515 (2020)}.

\bibitem{Dalianis2021}
I. Dalianis and K. Kritos,
\href{https://doi.org/10.1103/PhysRevD.103.023505}
{Phys. Rev. D \textbf{103}, 023505 (2021)}.

\bibitem{Yi2021}
Z. Yi, Y. Gong, B. Wang, and Z. Zhu,
\href{https://doi.org/10.1103/PhysRevD.103.063535}
{Phys. Rev. D \textbf{103}, 063535 (2021)}.

\bibitem{Gao2021}
Q. Gao, Y. Gong, and Z. Yi, 
\href{https://doi.org/10.1016/j.nuclphysb.2021.115480}
{Nucl. Phys. B \textbf{969}, 115480  (2021)}.

\bibitem{Yi2021b}
Z. Yi, Q. Gao, Y. Gong, and Z. Zhu,
\href{https://doi.org/10.1103/PhysRevD.103.063534}
{Phys. Rev. D \textbf{103}, 063534 (2021)}.

\bibitem{TGao2021}
T. Gao and X. Yang,
\href{https://doi.org/10.1140/epjc/s10052-021-09269-4}
{Eur. Phys. J. C \textbf{81}, 494 (2021)}. 

\bibitem{Solbi2021}
M. Solbi and K. Karami,
\href{https://doi.org/10.1088/1475-7516/2021/08/056}
{J. Cosmol. Astropart. Phys. \textbf{08} (2021) 056}.

\bibitem{Gao2021b}
Q. Gao,
\href{https://doi.org/10.1007/s11433-021-1708-9}
{Sci. China Phys. Mech. Astron. \textbf{64}, 280411 (2021)}.

\bibitem{Solbi2021b}
M. Solbi and K. Karami, 
\href{https://doi.org/10.1140/epjc/s10052-021-09690-9}
{Eur. Phys. J. C \textbf{81}, 884 (2021)}.

\bibitem{Zheng2021}
R. Zheng, J. Shi, and T. Qiu,
\href{https://doi.org/10.1088/1674-1137/ac42bd}
{Chin. Phys. C \textbf{46} 045103}.

\bibitem{Teimoori2021a}
Z. Teimoori, K. Rezazadeh, M. A. Rasheed, and K. Karami,
\href{https://doi.org/10.1088/1475-7516/2021/10/018}
{J. Cosmol. Astropart. Phys. \textbf{10} (2021) 018}.

\bibitem{Cai2021}
R. Cai, C. Chen,  and C. Fu,
\href{https://doi.org/10.1103/PhysRevD.104.083537}
{Phys. Rev. D \textbf{104},  083537 (2021)}.

\bibitem{Wang2021}
Q. Wang, Y. Liu, B. Su, and N. Li,
\href{https://doi.org/10.1103/PhysRevD.104.083546}
{Phys. Rev. D \textbf{104}, 083546 (2021)}.

\bibitem{Karam2022}
A. Karam, N. Koivunen, E. Tomberg, V. Vaskonen, and H. Veerm{\"a}e, 
\href{https://doi.org/10.48550/arXiv.2205.13540}{arXiv.2205.13540}.


\bibitem{Fuchengjie2019}
C. Fu, P. Wu, and H. Yu,
\href{https://doi.org/10.1103/PhysRevD.100.063532}
{Phys. Rev. D \textbf{100}, 063532 (2019)}.

\bibitem{fuchengjie2020}
C. Fu, P. Wu, and H. Yu,
\href{https://doi.org/10.1103/PhysRevD.101.023529}
{Phys. Rev. D \textbf{101}, 023529 (2020)}.

\bibitem{Dalianis(2020)}
I. Dalianis, S. Karydas, and E. Papantonopoulos, 
\href{https://doi.org/10.1088/1475-7516/2020/06/040}
{J. Cosmol. Astropart. Phys. \textbf{06} (2020) 040}.

\bibitem{Teimoori2021}
Z. Teimoori, K. Rezazadeh, and K. Karami,
\href{https://doi.org/10.3847/1538-4357/ac01cf}
{Astrophys. J. \textbf{915}, 118 (2021)}.

\bibitem{Heydari2022}
S. Heydari, and K. Karami, 
\href{https://doi.org/10.1140/epjc/s10052-022-10036-2}
{ Eur. Phys. J. C \textbf{82}, 83 (2022)}.

\bibitem{Heydari2022b}
S. Heydari, and K. Karami, 
\href{https://doi.org/10.1088/1475-7516/2022/03/033}
{J. Cosmol. Astropart. Phys. \textbf{03} (2022) 033}.

\bibitem{yfcai2018}
Y. F. Cai, X. Tong, D. G. Wang and S. F. Yan,
\href{https://doi.org/10.1103/PhysRevLett.121.081306}
{Phys. Rev. Lett. \textbf{121} (2018) 081306}.

\bibitem{yfcai2019}
Y. F. Cai, C. Chen, X. Tong, D. G. Wang and S. F. Yan,
\href{https://doi.org/10.1103/PhysRevD.100.043518}
{Phys. Rev. D \textbf{100} (2019) 043518}.

\bibitem{c.chen2019}
C. Chen and Y. F. Cai,
\href{https://doi.org/10.1088/1475-7516/2019/10/068}
{J. Cosmol. Astropart. Phys. \textbf{10} (2019) 068}.

\bibitem{c.chen2020}
C. Chen, X. H. Ma and Y. F. Cai,
\href{https://doi.org/10.1103/PhysRevD.102.063526}
{Phys. Rev. D \textbf{102} (2020) 063526}.


\bibitem{Addazi2022}
A. Addazi, S. Capozziello, and Q. Gan,
\href{https://arxiv.org/abs/2204.07668}
{arXiv: 2204.07668}

\bibitem{Cai2020}
R. Cai, Z. Guo, J. Liu, L. Liu, and X. Yang,
\href{https://doi.org/10.1088/1475-7516/2020/06/013}
{J. Cosmol. Astropart. Phys. \textbf {06} (2020) 013}




\bibitem{lisa}
P. Amaro-Seoane,  H. Audley, S. Babak, et al. (LISA),
\href{https://doi.org/10.48550/arXiv.1702.00786}
{arXiv:1702.00786}.

\bibitem{taiji}
W. R. Hu and Y. L. Wu,
\href{https://doi.org/10.1093/nsr/nwx116}
{Natl. Sci. Rev. \textbf{4}, 685 (2017)}.

\bibitem{tianqin}
J. Luo,  L. S. Chen, H. Z. Duan, et al. (TianQin),
\href{https://doi.org/10.1088/0264-9381/33/3/035010}
{Class. Quant. Grav. \textbf{33}, 035010 (2016)}.	

\bibitem{pta1}
R. D. Ferdman,  R. van Haasteren, C. G. Bassa, et al.,
\href{https://doi.org/10.1088/0264-9381/27/8/084014}
{Class. Quant. Grav. \textbf{27}, 084014 (2010)}.

\bibitem{pta2}
G. Hobbs,  A.  Archibald2, Z. Arzoumanian, et al.,
\href{https://doi.org/10.1088/0264-9381/27/8/084013}
{Class. Quant. Grav. \textbf{27}, 084013 (2010)}.

\bibitem{pta3}
M.  A. McLaughlin,	
\href{https://doi.org/10.1088/0264-9381/30/22/224008}
{Class. Quant. Grav. \textbf{30}, 224008 (2013)}.

\bibitem{pta4}
G. Hobbs, 
\href{https://doi.org/10.1088/0264-9381/30/22/224007}
{Class. Quant. Grav. \textbf{30}, 224007 (2013)}.

\bibitem{Cai2019}
R. G. Cai, S. Pi, and M. Sasaki,
\href{https://doi.org/10.1103/PhysRevLett.122.201101}
{Phys. Rev. Lett. \textbf{122}, 201101 (2019)}.


\bibitem{F.Zhang2022}
F. Zhang,
\href{https://doi.org/10.1103/PhysRevD.105.063539}
{Phys. Rev. D \textbf{105}, 063539 (2022)}.

\bibitem{fengge2020}
F.  Zhang,
\href{https://doi.org/10.1088/1475-7516/2021/04/045}
{J. Cosmol. Astropart. Phys. \textbf{04} (2021) 045}.

\bibitem{Zhang2021}
F. Zhang, J. Lin, and Y. Lu,
\href{https://doi.org/10.1103/PhysRevD.104.063515}
{ Phys. Rev. D \textbf{104}, 063515 (2021)}.

\bibitem{SamuelPassaglia2019}
S. Passaglia, W. Hu, and H. Motohashi,
\href{https://doi.org/10.1103/PhysRevD.99.043536}
{Phys. Rev. D \textbf{99}, 043536 (2019)}.

\bibitem{Chul-Moon Yoo2019}
C. M. Yoo, J. O. Gong, and S. Yokoyama,
\href{https://doi.org/10.1088/1475-7516/2019/09/033}
{J. Cosmol. Astropart. Phys. \textbf{09} (2019) 033}.

\bibitem{G.Franciolini2018}
G. Franciolini, A. Kehagias, S. Matarrese and A. Riotto,
\href{https://doi.org/10.1088/1475-7516/2018/03/016}
{J. Cosmol. Astropart. Phys. \textbf{03} (2018) 016}.

\bibitem{Matthew2022}
M. W. Davies, P. Carrilho, D. J. Mulryne,
\href{https://doi.org/10.1088/1475-7516/2022/06/019}
{J. Cosmol. Astropart. Phys. \textbf{06} (2022)  019}.

\bibitem{QingGuoHuang2013}
Q. G. Huang and Y. Wang,
\href{https://doi.org/10.1088/1475-7516/2013/06/035}
{J. Cosmol. Astropart. Phys. \textbf{06} (2013) 035}.

\bibitem{BravoRafael2018}
B. Rafael, M. Sander, P. G. A, Bastian,
\href{https://doi.org/10.1088/1475-7516/2018/05/025}
{J. Cosmol. Astropart. Phys. \textbf{05} (2018) 025}.

\bibitem{Cai2018}
Y. Cai, X. Chen, M. H. Namjoo, M. Sasaki, D. Wang,  and Z. Wang,
\href{https://doi.org/10.1088/1475-7516/2018/05/012}{J. Cosmol. Astropart. Phys. \textbf{05} (2018) 012}.

\bibitem{VicenteAtal2018}
V. Atal and C. Germani,
\href{https://doi.org/10.1016/j.dark.2019.100275}
{Phys. Dark Univ. \textbf{24},  1002757  (2019)}.

\bibitem{VicenteAtal2019}
V. Atal, J. Garriga and A. Marcos-Caballero
\href{https://doi.org/10.1088/1475-7516/2019/09/073}
{J. Cosmol. Astropart. Phys. \textbf{09} (2019) 073}.

\bibitem{Germani2011_1} C. Germani and A. Kehagias, 
\href{https://doi.org/10.1103/PhysRevLett.106.161302}{Phys. Rev. Lett. \textbf{106}, 161302 (2011)}.

\bibitem{Germani2011_2}
 C. Germani and Y. Watanabe, 
\href{https://doi.org/10.1088/1475-7516/2011/07/031}
{J. Cosmol. Astropart. Phys. \textbf{07} (2011) 031}.

\bibitem{A.D.Felice2011}	
A. D. Felice and S.Tsujikawa,
\href{https://doi.org/10.1088/1475-7516/2011/04/029}	
{J. Cosmol. Astropart. Phys. \textbf{04} (2011) 029}.

\bibitem{S.Tsujikawa2012}	
S. Tsujikawa,
\href{https://doi.org/10.1103/PhysRevD.85.083518}	
{Phys. Rev. D \textbf{85}, 083518 (2012)}.

\bibitem{Kobayashi2011}
T. Kobayashi, M. Yamaguchi, and J. Yokoyama,		
\href{https://doi.org/10.1143/PTP.126.511}
{Prog. Theor. Phys. \textbf{126}, 511 (2011)}.

\bibitem{Silverstein2008}
E. Silverstein and A. Westphal,
\href{https://doi.org/10.1103/PhysRevD.78.106003}
{Phys. Rev. D \textbf{78}, 106003 (2008)}.

\bibitem{Ade2021} P. A. R. Ade, Z. Ahmed, M. Amiri, et al.,
\href{https://doi.org/10.1103/PhysRevLett.127.151301}
{Phys. Rev. Lett. \textbf{127}, 151301 (2021)}.

\bibitem{C.T.Byrnes2010}
C. T. Byrnes, M. Gerstenlauer, S. Nurmi, G. Tasinato, and D. Wands,
\href{https://doi.org/10.1088/1475-7516/2010/10/004}
{J. Cosmol. Astropart. Phys. \textbf{10} (2010) 004}.


\bibitem{P.Ade2016}
P. A. R. Ade, N. Aghanim, M. Arnaud, et al,
\href{https://doi.org/10.1051/0004-6361/201525836}
{Astron. Astrophys. \textbf{594}, A17 (2016)}.

\bibitem{DavidSeery2005}
D. Seery and J. E. Lidsey,
\href{https://doi.org/10.1088/1475-7516/2005/06/003}
{J. Cosmol. Astropart. Phys. \textbf{06} (2005) 003}.

\bibitem{X.Chen2007}
X. Chen, R. Easther, and E. A. Lim,
\href{https://doi.org/10.1088/1475-7516/2007/06/023}
{Cosmol. Astropart. Phys. \textbf{06} (2007) 023}.

\bibitem{FredericoArroja2011}
F. Arroja and T. Tanaka,
\href{https://doi.org/10.1088/1475-7516/2011/05/005}
{J. Cosmol. Astropart. Phys. \textbf{05} (2011) 005}.

\bibitem{Maldacena2003}
J. M. Maldacena,
\href{https://doi.org/10.1088/1126-6708/2003/05/013}
{J. High Energ. Phys. \textbf{05} (2003) 013}.


\bibitem{Creminelli2002}
P. Creminelli, L. Senatore, M. Zaldarriaga, and M. Tegmark,
\href{https://doi.org/10.1088/1475-7516/2007/03/005}
{J. Cosmol. Astropart. Phys. \textbf{03} (2007) 005}.

\bibitem{X.Chen2008}
X. Chen, R. Easther, and E. A. Lim,
\href{https://doi.org/10.1088/1475-7516/2008/04/010}
{J. Cosmol. Astropart. Phys. \textbf{04} (2008) 010}.

\bibitem{Tada2019}
Y. Tada and S. Yokoyama,
\href{https://doi.org/10.1103/PhysRevD.100.023537}
{Phys. Rev. D \textbf{100}, 023537 (2019)}.

\bibitem{Young2014}
S. Young, C. T. Byrnes, and M. Sasaki,
\href{https://doi.org/10.1088/1475-7516/2014/07/045}	
{J. Cosmol. Astropart. Phys. \textbf{07} (2014) 045}.

\bibitem{Musco2013}
I. Musco and J. C. Miller,
\href{https://doi.org/10.1088/0264-9381/30/14/145009}	
{Class. Quant. Grav. \textbf{30}, 145009 (2013)}.

\bibitem{Harada2013}
T. Harada, C. M. Yoo, and K. Kohri,
\href{https://doi.org/10.1103/PhysRevD.88.084051}
{Phys. Rev. D \textbf{88}, 084051 (2013)}.


\bibitem{Riccardi2021}
F. Riccardi, M. Taoso, and A. Urbano,
\href{https://doi.org/10.1088/1475-7516/2021/08/060}
{J. Cosmol. Astropart. Phys. \textbf{08} (2021) 060}.

\bibitem{Inomata2019}
K. Inomata and T. Nakama,
\href{https://doi.org/10.1103/PhysRevD.99.043511}
{Phys. Rev. D \textbf{99}, 043511  (2019)}.

\bibitem{Kohri2018}
K. Kohri and T. Terada,
\href{https://doi.org/10.1103/PhysRevD.97.123532}
{Phys. Rev. D \textbf{97}, 123532 (2018)}.

\bibitem{Verde2000}
L. Verde, L. M. Wang, A. Heavens, and M. Kamionkowski,
\href{https://doi.org/10.1046/j.1365-8711.2000.03191.x}
{Mon. Not. Roy. Astron. Soc. \textbf{313}, L141 (2000)}.

\bibitem{Komatsu2001}
E. Komatsu and D. N. Spergel,
\href{https://doi.org/10.1103/PhysRevD.63.063002}
{Phys. Rev. D \textbf{63}, 063002 (2001)}.


\bibitem{Felice2011}
A. De Felice and S. Tsujikawa,
\href{https://doi.org/10.1103/PhysRevD.84.083504}
{Phys. Rev. D \textbf{84}, 083504 (2011)}.



\end{thebibliography}
\end{document}